\documentclass[11pt]{article}
\usepackage{moriond,epsfig}

\def\Journal#1#2#3#4{{#1} {\bf #2}, #3 (#4)}

\def\PLB{{\em Phys. Lett.}  B}
\def\PRL{\em Phys. Rev. Lett.}
\def\PRD{{\em Phys. Rev.} D}

\def\be{\begin{equation}}
\def\ee{\end{equation}}
\def\bea{\begin{eqnarray}}
\def\eea{\end{eqnarray}}

\begin{document}
\vspace*{4cm}
\title{$W/Z$+jet results from the Tevatron}

\author{D.V.~Bandurin (for the D0 and CDF Collaborations)}

\address{Department of Physics, Florida State University, Tallahassee, Florida 32306, USA}

\maketitle
\abstracts{
Selected quantum chromodynamics (QCD) measurements performed at the Fermilab
Run II Tevatron $p\bar{p}$ collider running at $\sqrt{s}=1.96$ TeV by D0 and CDF
Collaborations are presented.
Events with $W/Z$+jets productions are used to measure
many kinematic distributions allowing extensive tests and tunes of predictions from 
perturbative QCD at next-to-leading (NLO) order and Monte-Carlo (MC) event generators.
}

\section{Introduction}
The D0 and CDF collaborations have extensively studied the $W/Z$+jet productions since these
events are the main background to top-quark, Higgs boson, SUSY and many other new
physics production channels.
To make discoveries at the Tevatron and the LHC,
these processes need to be measured and simulated with a level of accuracy that will be comparable 
to the significance of the new physics signals.

There are several programs on the market that
can simulate hadronic interactions at NLO accuracy, but the processes included in
these programs are limited. Matrix element plus parton shower (ME+PS) programs 
simulate a more comprehensive set of processes, typically at leading-log
(LL) or leading order (LO) accuracy, and rely on models to
simulate emissions and fragmentation associated with
higher order processes. These programs have been
employed regularly for background simulation at the
Tevatron in recent years, notably in the Higgs searches~\cite{HW} 
and the discovery of the production of single top quarks \cite{SingleTop}.
The Tevatron measurements presented here are compared to
predictions by NLO pQCD in MCFM~\cite{MCFM}, BLACKHAT+SHERPA~\cite{BS} and 
ROCKET+MCFM~\cite{RM},  ME+PS programs ALPGEN \cite{Alpgen} 
and SHERPA \cite{Sherpa}, and PS programs HERWIG \cite{Herwig} and PYTHIA \cite{Pythia}. 
The most of measurements have been published \cite{Zjets_D0,Wjets_D0,Zb_D0,Wc_D0,Wb_CDF} 
or approved as preliminary results  \cite{Zjets_CDF,Wc_CDF,Zb_CDF} at the time 
these proceedings were written. ALPGEN employs the MLM
algorithm to ensure jets originating from the matrix element and the parton shower are not double
counted. SHERPA is a CKKW-inspired model which
uses a re-weighting of the matrix elements to achieve
the same appropriate jet configurations. A detailed
description of these programs can be found in \cite{MEPS}.

In this paper we review some of the recent Tevatron results on the  $W/Z$+jet and
$W/Z$+heavy flavor jet productions.

\subsection{$W/Z$+jet production}\label{subsec:WZ_Jets}

Both collaborations have extensively studied the $W/Z$+jet productions 
with $Z$ and $W$ decaying via electron and muon decay modes.
The leptonic decay of the $Z/W$ provides a clean signal for reconstruction 
of the events, and small background contamination.
The test of pQCD is made by comparing the measurements to NLO pQCD
predictions. The $W/Z$ + jets final states also make up a major 
background of many new physics searches at both the
Tevatron and LHC. Therefore, these data measurements unfolded to the particle 
level are useful for tuning LO simulation programs which are heavily relied
upon to model background processes.

\begin{figure}[htbp]
\hspace*{5mm}  \includegraphics[scale=0.17]{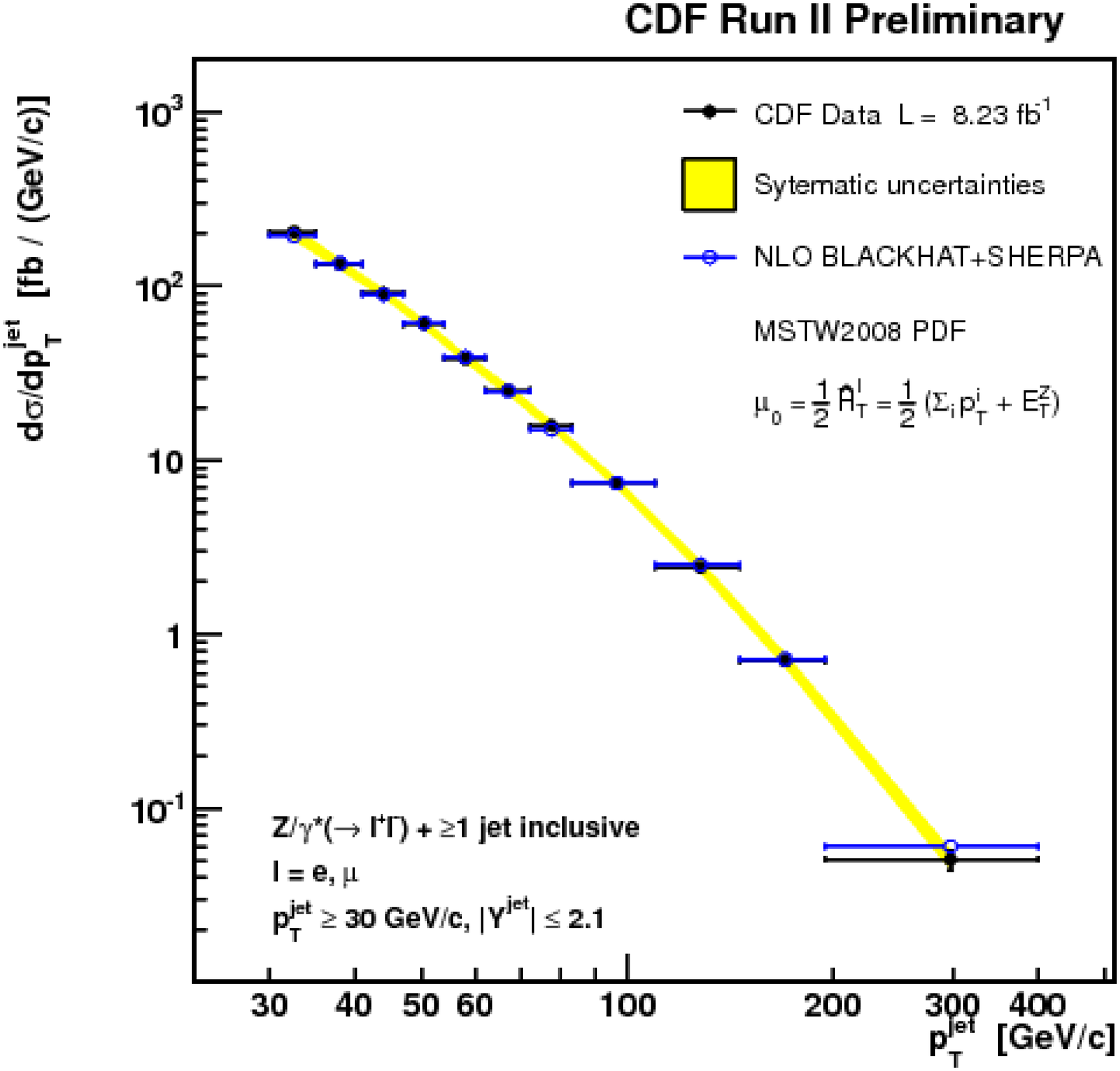}
\hspace*{10mm}  \includegraphics[scale=0.17]{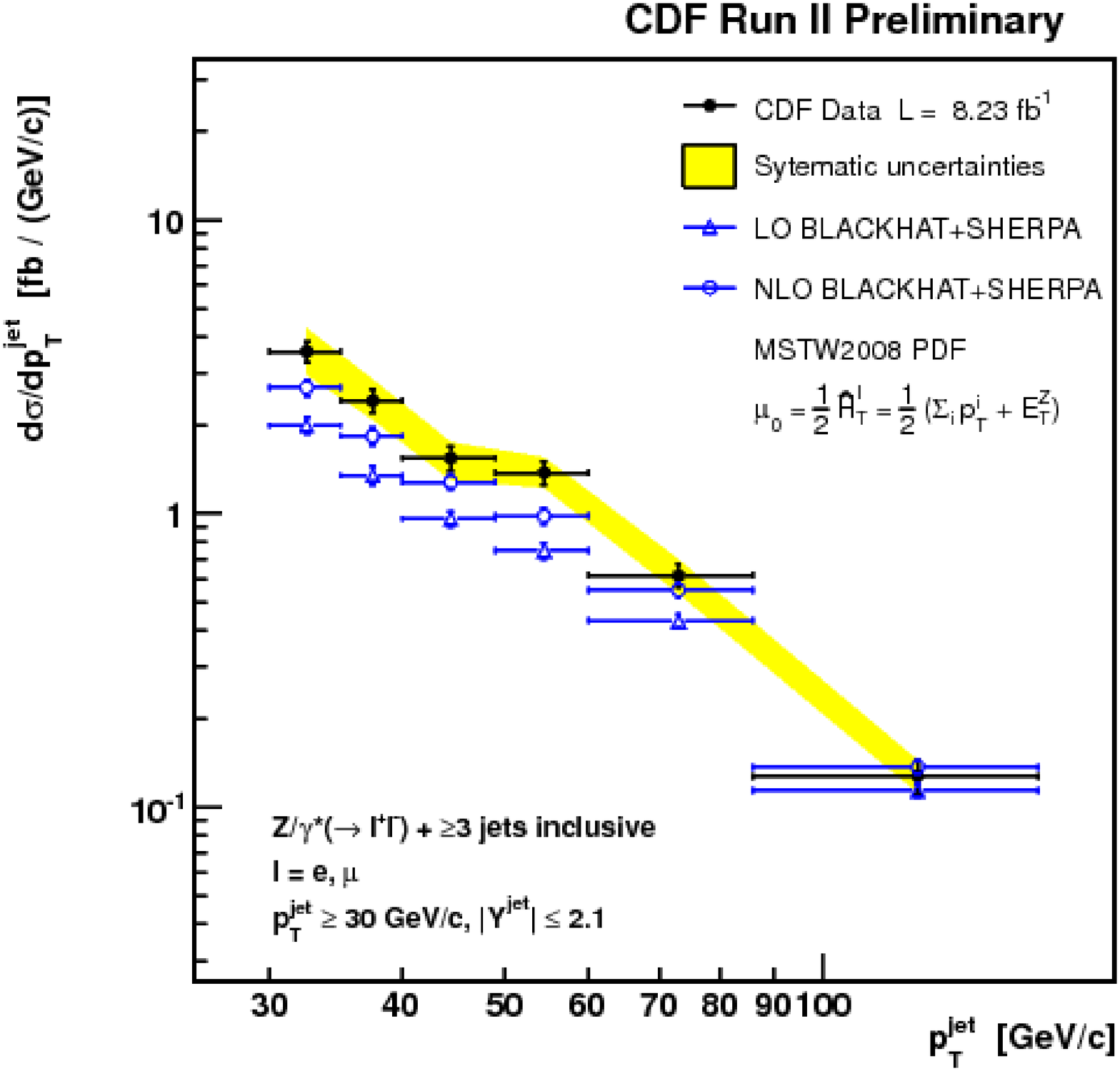}  

\hspace*{5mm}  \includegraphics[scale=0.17]{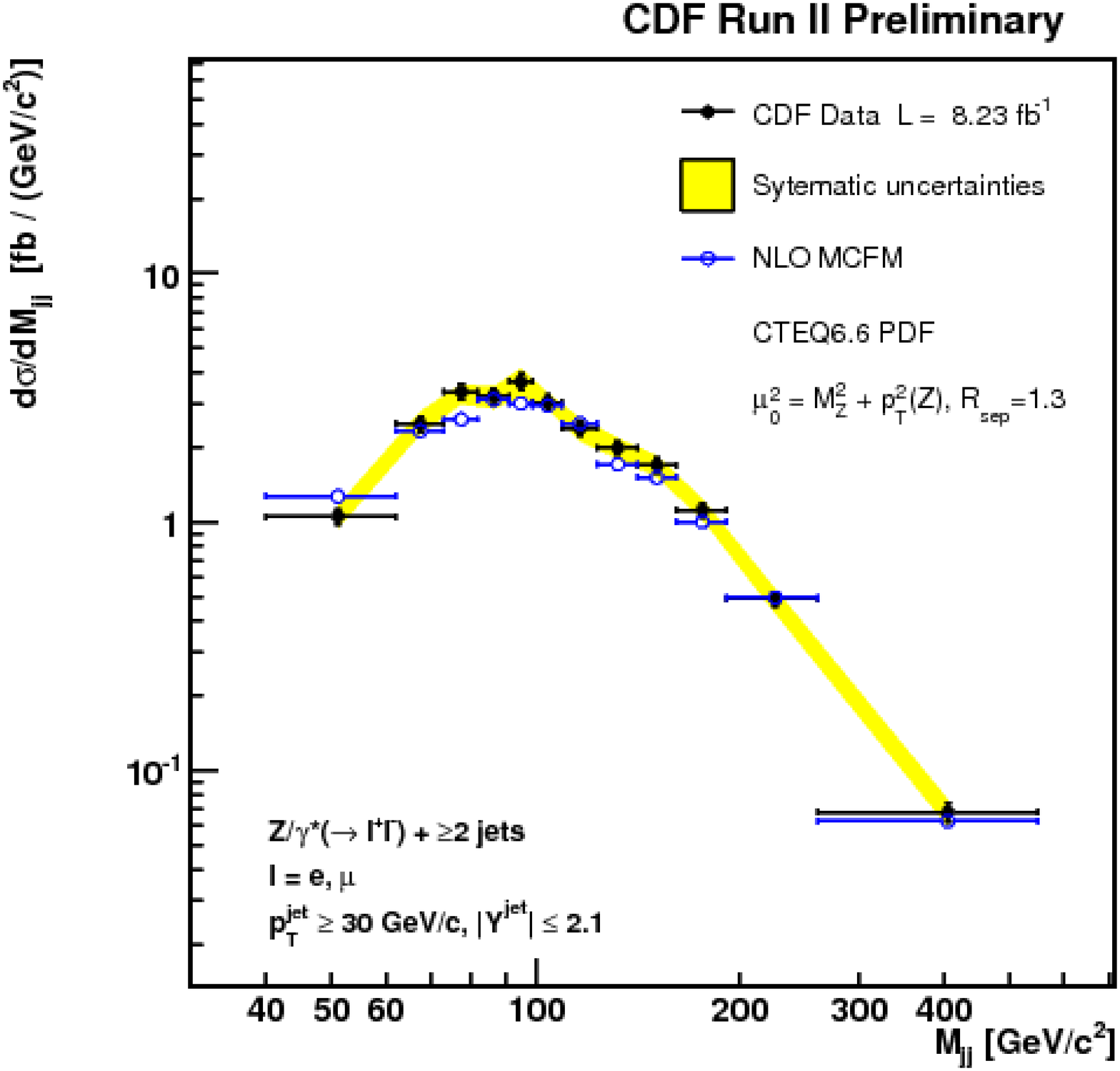}
\hspace*{10mm}  \includegraphics[scale=0.17]{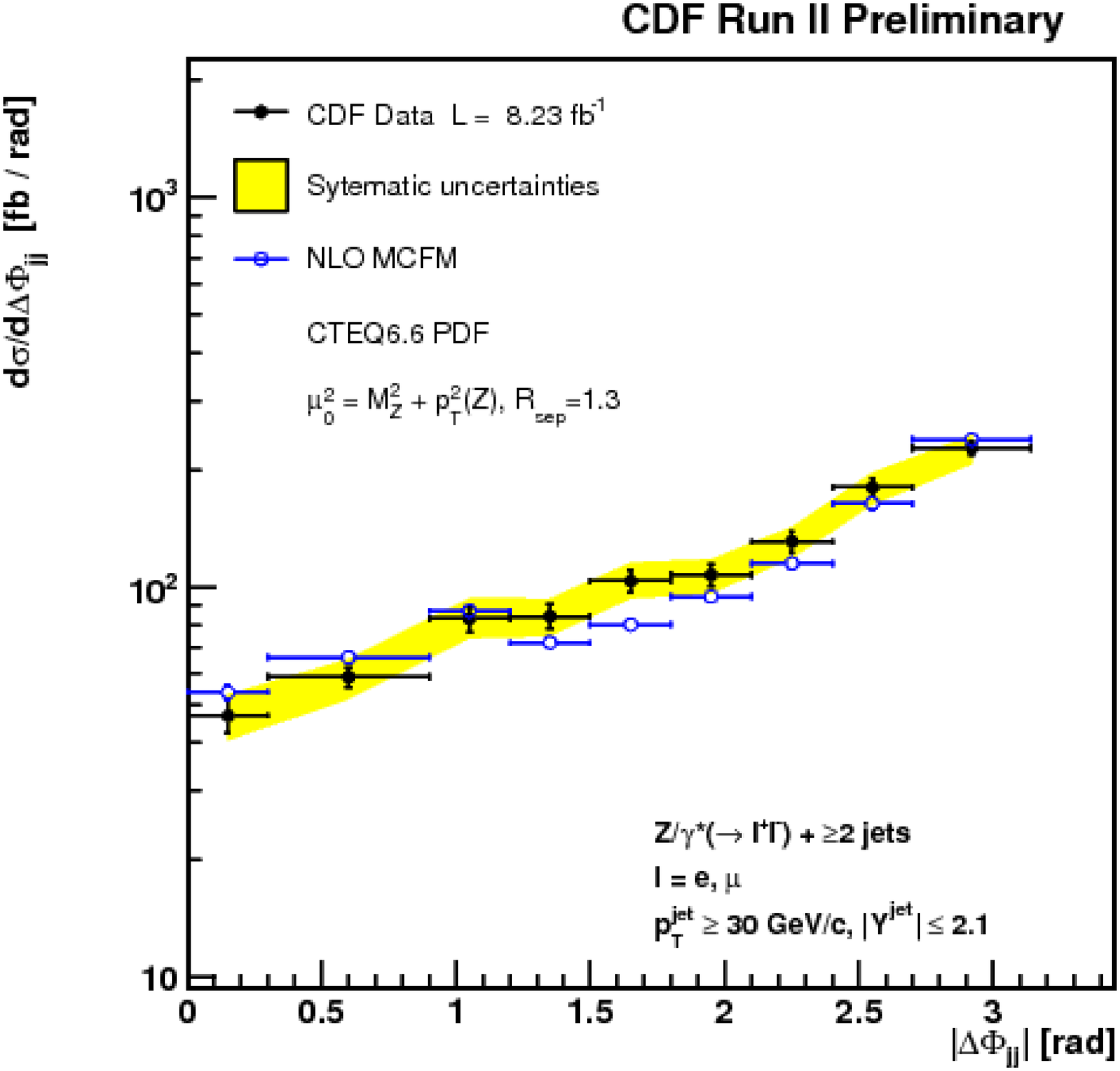}
\caption{Two top plots show measured inclusive cross section for $Z/\gamma^*$+jets production as a function 
of leading jet $p_T$ in $Z+\!\geq\!\!1$ jet events (top left) and 3rd jet $p_T$ in $Z+\!\geq\!\!3$ jet events (top right)
compared to NLO pQCD predictions using BLACKHAT+SHERPA. 
Two bottom plots show measured cross section as a function of 
dijet invariant mass (bottom left) and azimuthal angle between two jets (bottom left) in $Z+\!\geq\!\!2$ jet events;
results are compared to NLO pQCD predictions using MCFM.}
\label{fig:zj_cdf}
\end{figure}

Fig.~\ref{fig:zj_cdf} shows the inclusive cross section for $Z/\gamma^*$+jets production measured
by CDF \cite{Zjets_CDF} as a function of leading and 3rd jet $p_T$ (jets are ordered in descending $p_T$) 
in $Z+\!\geq\!\!1$ jet and in $Z\!+\geq\!\!3$ jet events. Also shown are 
dijet invariant mass and azimuthal angle between the two leading jets in $Z+\geq2$ jet events.
The measurements are in agreement with NLO pQCD predictions (BLACKHAT+SHERPA and MCFM)
within theoretical scale uncertainties which are about 25\%, obtained
by variation of the default scale by a factor 2.

D0 measured jet $p_T$ inclusive cross sections of $W+n$-jet production for jet multiplicities $n=1-4$ 
\cite{Wjets_D0}.
The measurements are compared to the NLO predictions for $n=1-3$
and to LO predictions for $n = 4$. The measured cross sections are generally found to
agree with the NLO calculation although certain regions
of phase space are identified where the calculations could be improved.

\begin{figure}[htbp]
\vspace*{-10mm}
\hspace*{0mm}  \includegraphics[scale=0.30]{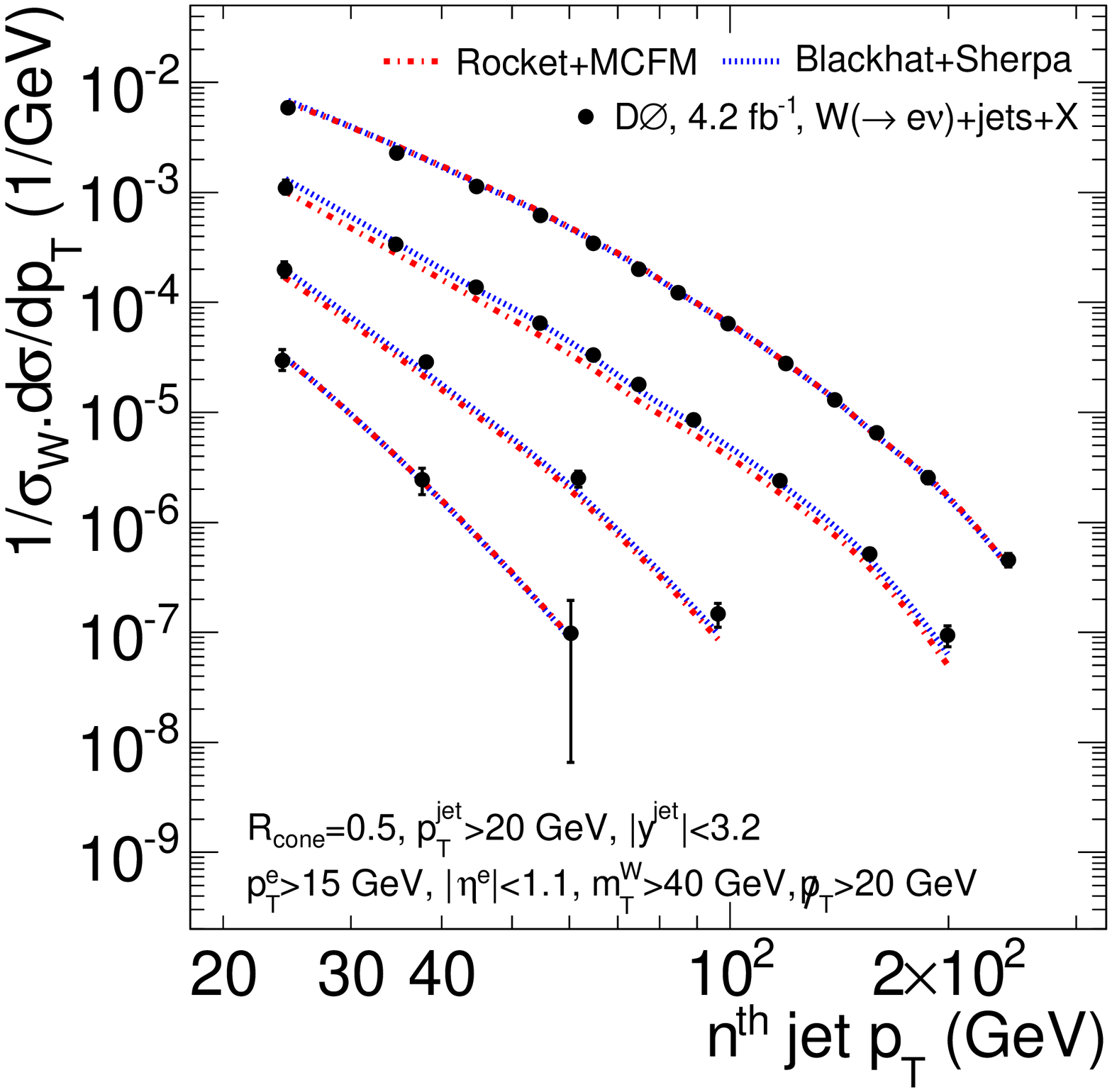}
\hspace*{7mm}  \includegraphics[scale=0.28]{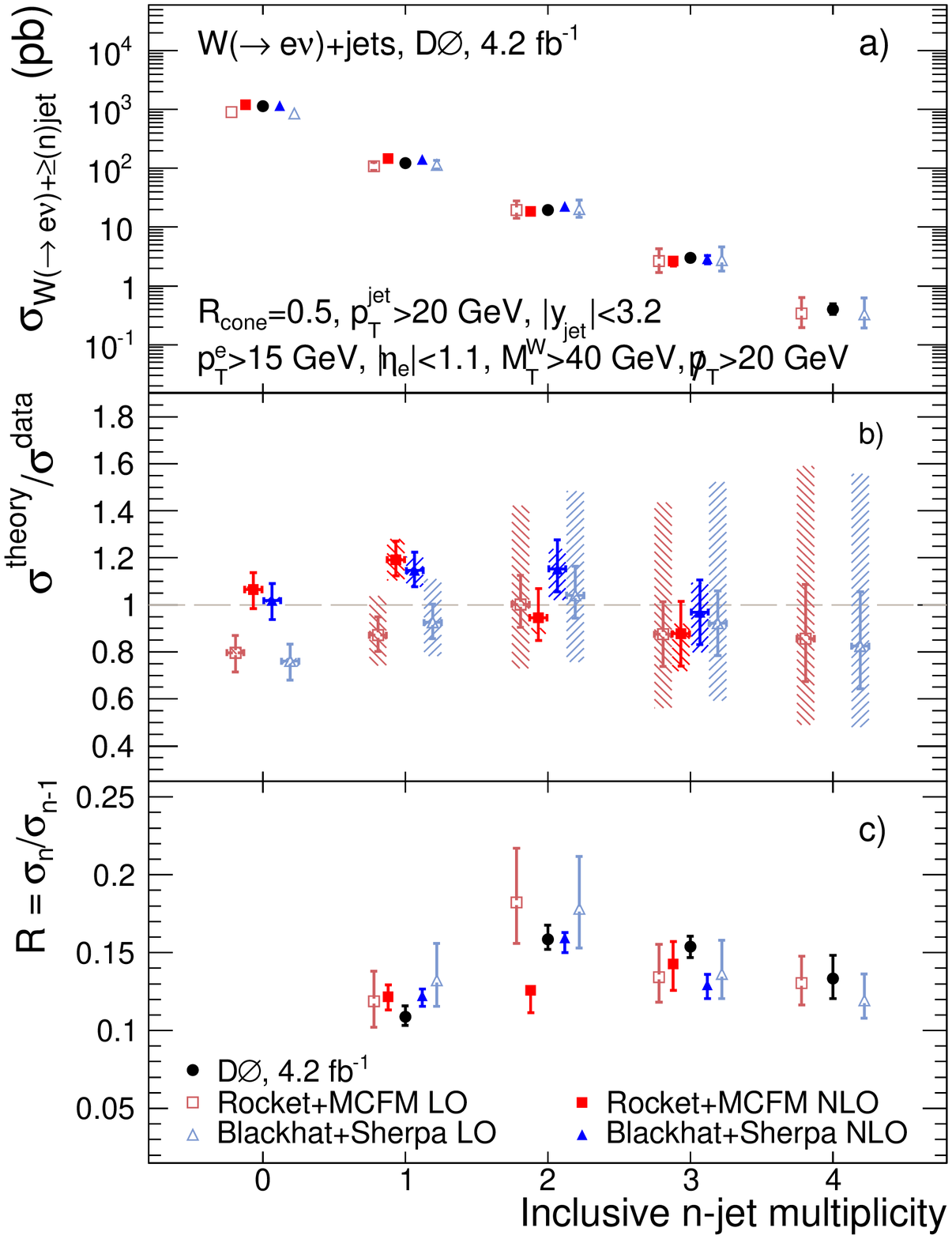}
\caption{Left: measured $W +n$ jet differential cross section as a function of jet 
$p_T$ for $n=1-4$, normalized to the inclusive $W\to e\nu$ cross section. The $W +1$ jet inclusive spectra
are shown by the top curve, the $W +4$ jet inclusive spectra by
the bottom curve. The measurements are compared to the
fixed-order NLO predictions for $n=1-3$ and to LO predictions for $n = 4$.
Right: (a) total inclusive $n$-jet cross sections 
$\sigma_n$ 
as a function of $n$, (b) the ratio of the theory predictions to the measurements, 
and (c) $\sigma_n/\sigma_{n-1}$ ratios for data,
Blackhat+Sherpa and Rocket+MCFM. 
The hashed areas represent the theoretical uncertainty arising from the choice of
renormalization and factorization scale.}
\label{fig:wj_d0}
\vspace*{-2mm}
\end{figure}

\subsection{$W/Z$+heavy flavor jet production}
\label{subsec:WZ_HF}

\begin{figure}[htbp]
\vspace*{-5mm}
\hspace*{10mm}  \includegraphics[scale=0.24]{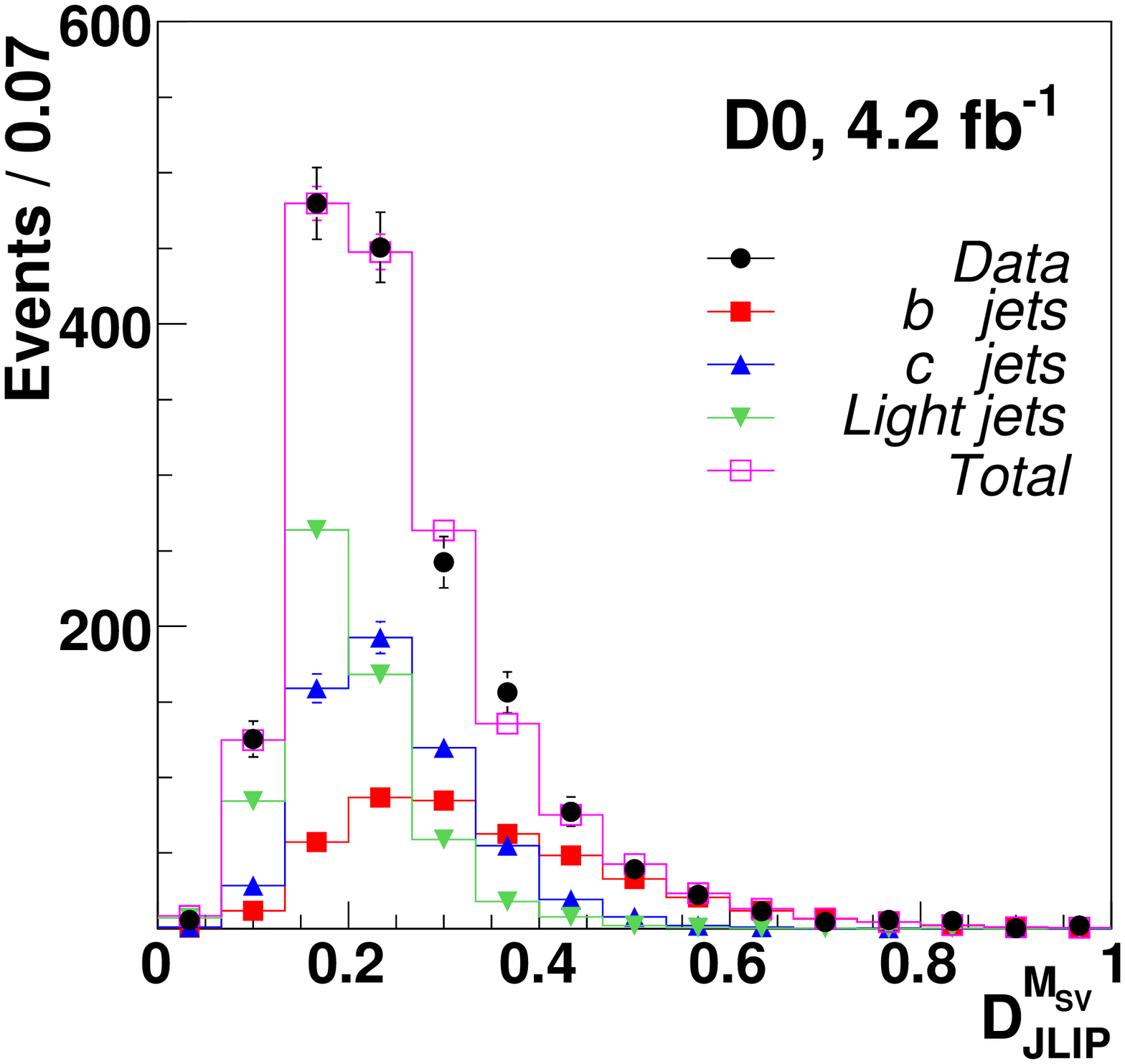}
{\vspace*{5mm}
\hspace*{19mm}  \includegraphics[scale=0.25]{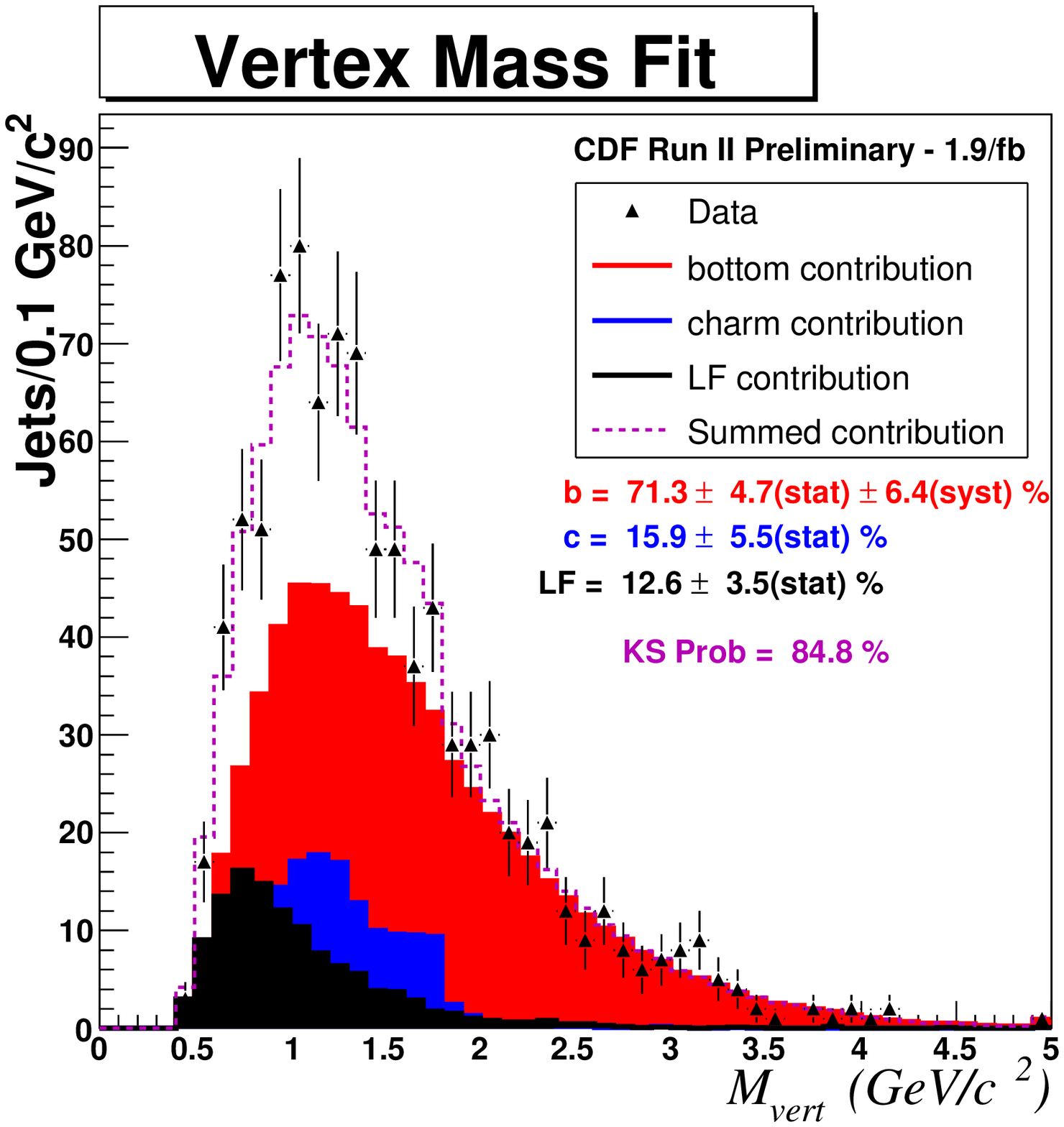}}
\vspace*{-9mm}
\caption{Left (D0):  
the distributions of the $b$, $c$, light jets and data over the $b-$jet discriminant;
MC templates are weighted by the fractions found from the fit to data. 
Right (CDF): the secondary vertex mass fit for the tagged jets in the selected sample.} 
\label{fig:wzb}
\vspace*{-5mm}
\end{figure}

D0 recently published the measured cross section ratio 
$\sigma(Z+b)/\sigma(Z+{\rm jet})=0.0193\pm0.0022({\rm stat})\pm0.0015({\rm syst})$
for events with jet $p_T>20$ GeV and $|\eta|<2.5$ \cite{Zb_D0}. 
This most precise measurement of the $Z+b$ fraction is consistent with the NLO theory prediction, $0.0192\pm0.0022$,
done with MCFM, renormalization and factorization scales set at $m_Z$, and 
the CDF result $0.0208\pm0.0033({\rm stat})\pm0.0034({\rm syst})$ \cite{zb_CDF}.
The CDF collaboration measured the cross section of $W+b$-jet production 
$\sigma(W+b)\cdot Br(W\to l\nu) = 2.74\pm 0.27({\rm stat})\pm0.42({\rm syst})$ pb
with jet $p_T>20$ GeV, $|\eta|<2.0$ and $l=e,\mu$. 
The measurement significantly exceeds the NLO prediction $1.2\pm0.14$ pb.
The fit results for the $b$-jet fractions for both the measurements are shown in Fig.~\ref{fig:wzb}. \\[-5mm]

\begin{figure}[htbp]
\hspace*{5mm} \includegraphics[scale=0.18]{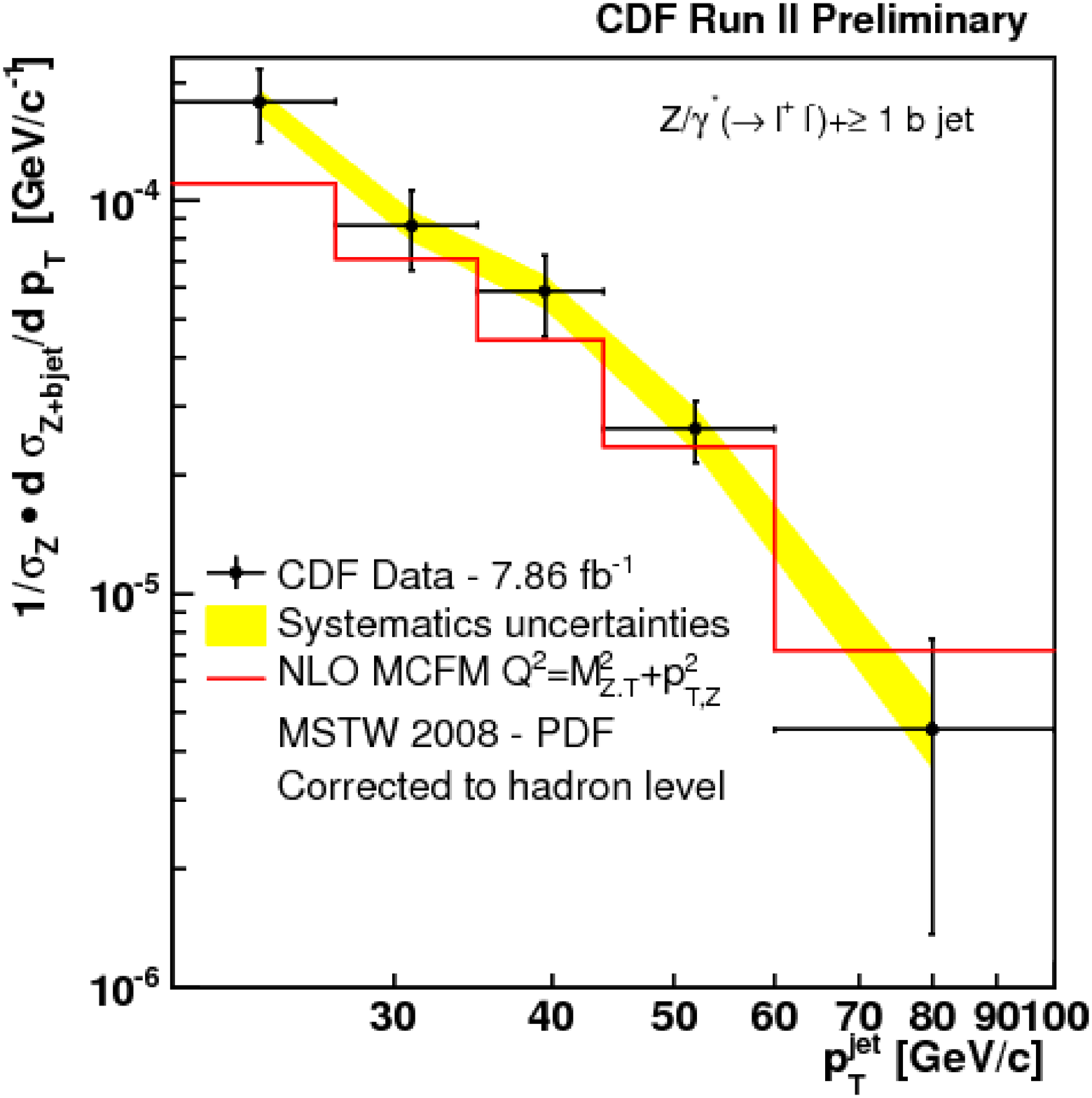}
\hspace*{10mm} \includegraphics[scale=0.18]{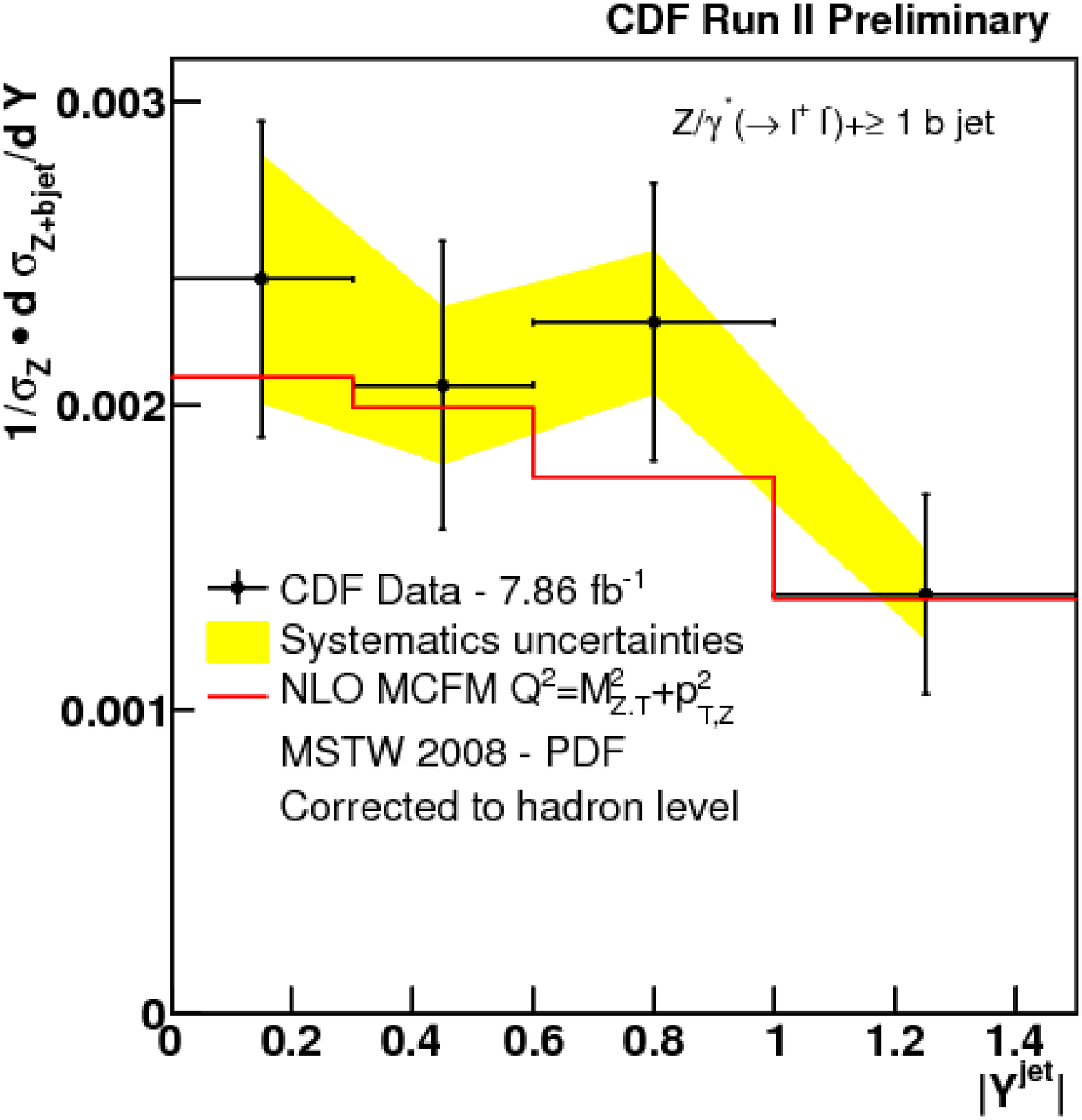}
\vspace*{-4mm}
\caption{Differential cross section of $Z+b$ production as a function
of $b$-jet $p_T$ (left) and rapidity (right).}
\label{fig:zb_cdf}
\vspace*{-3mm}
\end{figure}

The CDF collaboration has also measured differential cross section of $Z+b$-jet production 
versus $b$-jet $p_T$ and $\eta$~\cite{Zb_CDF}. 
Results are shown in Fig.~\ref{fig:zb_cdf}.
The following cross sections ratio have been also measured,
$\sigma(Z+b)/\sigma(Z) = 0.293\pm0.030({\rm stat})\pm0.036({\rm syst})\%$ and
$\sigma(Z+b)/\sigma(Z+jet) = 2.31\pm0.23({\rm stat})\pm0.32({\rm syst})\%$. 

Both experiments measured $W+c$ production cross section using the soft lepton tagging technique
\cite{Wc_D0,Wc_CDF}. The D0 collaboration measured ratio $\sigma(W+c)/\sigma(W+jet)$ and
found it to be $0.074\pm0.019({\rm stat})^{+0.012}_{-0.014}({\rm syst})\%$, what is higher than 
ALPGEN+PYTHIA predictions $0.044\pm0.003$.
The CDF collaboration measured total cross section (electron and muon channels combined)
and found  $\sigma(W+c)\times Br(W\to l\nu, l=e,\mu) = 13.3^{+3.3}_{-2.9}$ pb what
is in agreement with pQCD NLO predictions $11.3\pm 2.2$ pb.

\section*{Summary}
Several differential cross sections of $W/Z$ + jet$ + X$
events measured with the D0 and CDF detectors have been 
presented. The data are generally consistent with predictions 
from NLO pQCD, although some LO programs can also reproduce the shape 
of the data, sometimes better than NLO,
due either to their inclusion of higher parton multiplicity matrix elements 
than can be currently included in a fixed order pQCD calculation, or an optimized tune
of MC. These data should be useful for
continued tuning of these and other MC programs used at the Tevatron and LHC experiments.

\section*{Acknowledgments}
The author would like to thank Mike Strauss and Christina Mesropian
for a help in preparing the conference talk and this paper.

\end{document}